\documentclass[prb,letterpaper,floatfix,nobalancelastpage,twocolumn,superscriptaddress]{revtex4}
\usepackage{dcolumn}
\usepackage{bm}
\usepackage{amsmath}
\usepackage{bm,braket}
\usepackage{color}
\usepackage{float}
\usepackage{graphicx,subfigure}
\usepackage{overpic}
\usepackage{amssymb}
\usepackage{sidecap}

\definecolor{mmagenta}{RGB}{255, 0, 239}
\newcommand{\rem}[1]{}

\begin{document}

\title{Floquet Topological Polaritons in Semiconductor Microcavities}

\author{R. Ge}
\affiliation{Division of Physics and Applied Physics, School of Physical and Mathematical Sciences, Nanyang Technological University, 21 Nanyang Link, Singapore 637371}

\author{W. Broer}
\affiliation{Division of Physics and Applied Physics, School of Physical and Mathematical Sciences, Nanyang Technological University, 21 Nanyang Link, Singapore 637371}

\author{T. C. H. Liew}
\affiliation{Division of Physics and Applied Physics, School of Physical and Mathematical Sciences, Nanyang Technological University, 21 Nanyang Link, Singapore 637371}

\date{\today}

\begin{abstract}
We propose and model Floquet topological polaritons in semiconductor microcavities, using the interference of frequency detuned coherent fields to provide a time periodic potential. For arbitrarily weak field strength, where the Floquet frequency is larger than the relevant bandwidth of the system, a Chern insulator is obtained. As the field strength is increased, a topological phase transition is observed with an unpaired Dirac cone proclaiming the anomalous Floquet topological insulator. As the relevant bandwidth increases even further, an exotic Chern insulator with flat band is observed with unpaired Dirac cone at the second critical point. Considering the polariton spin degree of freedom, we find that the choice of field polarization allows oppositely polarized polaritons to either co-propagate or counter-propagate in chiral edge states.
\end{abstract}

\maketitle
\section{introduction}
When the bands of a material are transformed in a discontinuous way, that is inequivalent to any continuous transformation in the adiabatic limit, it becomes topologically distinct from its surroundings. As a consequence, gapless states emerge at the edges of such a material, which may have remarkable properties. For example, topological insulators~\cite{topo1,topo2} may have a bulk bandgap yet support chiral edge states propagating only in a specific direction. Aside emphasizing the relevance of characterizing phases of materials by the possibility of transforming them or not into one another (that is, their possible homeomorphism), the resulting dissipationless channels are relevant for signal/energy transmission and collection. The engineering of the non-trivial twist in the bandstructure to break homeomorphism with regular materials relies on a gauge field (usually a magnetic field). In electronic topological insulators the gauge field is responsible for coupling spin and orbital motion~\cite{tpreview1,tpreview2,QSH,Z21}, but it is hard to find materials with strong spin-orbit coupling. Recently, there has been an increasing interest in simulating/achieving topological insulator analogues in acoustics~\cite{acoutop}, (nano)photonics~\cite{topphot0,topphot01,topphot1,topphot2,topphot3,topphot4,photto1}, (nano)plasmonics~\cite{topplas1,topplas2}, ultracold atoms~\cite{atom1,Z2atom,atomr} and exciton-polaritons~\cite{topopolar1,topopolar2,topopolar3,topopolar4,topopolar5,topopolar6} to name a few.

Many non-electronic systems tend to operate with neutral particles, which do not experience an effect of magnetic field on orbital motion, so it is necessary to engineer artificial gauge fields. Here Floquet toplogical insulators~\cite{ftopo1,ftopochar1} have emerged, which use a time-periodic potential to force particles along the same cyclotron-type trajectories as a magnetic field operating on electrons. There has been a tremendous interest in exploring the topological phases with Floquet manipulation starting from topologically trivial states~\cite{photfto1,ftopochar2,ftopola,FBT,photfto2}.

As an efficient way of coherent manipulation by time-periodic forcing, shaken optical lattices have been employed with ultracold atoms to achieve artificial gauge structures and topologically non-trivial phases~\cite{magnet,nonabe,floqrpp,anyon,Floqrmp}; and the quantum anomalous Hall effect has been realized recently~\cite{Floqzhai,atomano}. However, the typical temperature required to achieve quantum degenerate gases in experiment is extremely low and the relevant optical lattice constant employed is at sub-micrometer scale, making experiments highly challenging. In contrast, the quantum coherence of exciton-polaritons in semiconductor microcavities can survive even at room temperature~\cite{roomt1,roomt2,roomt3,roomt4,roompolaritons} and with length scale as large as one millimeter~\cite{millimeter}. These merits, together with the versatile experimental techniques well developed to explore the phase coherence and spectrum of the exciton-polaritons both in real and reciprocal spaces, suggest that exciton-polaritons can complement the ultracold atomic system to explore nontrivial topological effects.

Several proposals of topological polaritons~\cite{topopolar1,topopolar2,topopolar4,topopolar5,topopolar6}, have relied on spin-orbit coupling via transverse electric-transverse magnetic (TE-TM) splitting of the polariton modes. Since this is limited in many samples, alternative mechanisms of artificial gauge fields for polaritons are important~\cite{magpolariton}. Remarkably, Floquet control of polaritons has not been considered in spite of its successful application with ultracold atoms. Unlike ultracold atoms, exciton-polaritons are (typically) a highly non-equilibrium systems employed to study the interplay between Bose-Einstein condensation and drive/dissipation, low threshold lasing, superfluidity and other effects (see, e.g.,~\cite{polaritonreview1,polaritonreview2} for a review). The achievement of (Floquet) topological states for exciton-polaritons could open a door to study the interplay between topological phase, interactions~\cite{topopolar3,interacting} and boson statistics.

The extended Hilbert space of a time-periodic Hermitian system ($H(t)$), with period $T$,  separates into branches with width given by the Floquet frequency, $\omega=2\pi/T$. If the energy scale of the Hermitian system at any time point, $0\le t <T$, is much smaller than the Floquet frequency, $\max(E_i(t))-\min(E_j(t))\ll \omega$, then higher Floquet bands could be adiabatically eliminated giving an effective Hamiltonian description~\cite{API}, $H_{\rm eff}$ (Magnus/High Frequency Expansion~\cite{HFE1,HFE2,HFE3}) and the topological classification is given by that of the well-studied static systems~\cite{topochar1,topochar2,topochar3}. On the other hand, if the energy scale of the system is comparable or even larger than the Floquet frequency, different branches are coupled closely and the idea of eliminating higher Floquet bands does not work. In this case, a faithful description requires keeping all the branches which have nonzero overlap in some spectral range $[\min(E_i(t)), \max(E_j(t))]$. Phenomenologically, this could lead to the case where the quasi-energy band of the Floquet operator $U(T) = \mathcal{T}\exp(-i\int_0^TH(t)dt/\hbar)$ is unbounded in contrast to the static case; as a result anomalous edge states can appear even as the Chern numbers ($\mathcal{C}$) of the bands of $H_{\rm eff}$ are zero. In the following, we first introduce our model of shaken lattice for the exciton-polaritons in Sec.~\ref{S:model}, and the simulation technique is presented. In Sec.~\ref{S:result}, we show that both the Chern insulator (with $\mathcal{C}=\pm1$) and the anomalous Floquet topological insulator (with $\mathcal{C}=0$, yet topological edge states) are possible in the exciton-polariton system. We will also show that, when the bandwidth is increased even further, an exotic Chern insulator appears, characterized by $\mathcal{C}=\pm1$ and an additional flat band. We compare calculations from continuous potential and tight-binding calculations (as shown in the Appendix). The latter show that large Chern insulators, with $\mathcal{C}=\pm5$ can be obtained when the bandwidth increases further with respect to the Floquet frequency. The summary is given in Sec.~\ref{S:3}.

\section{Model and theory}
\label{S:model}
\begin{figure}
\centering
\setbox1=\hbox{\includegraphics[trim=1.8cm 0.1cm 1.1cm .1cm, clip=true,height=8cm]{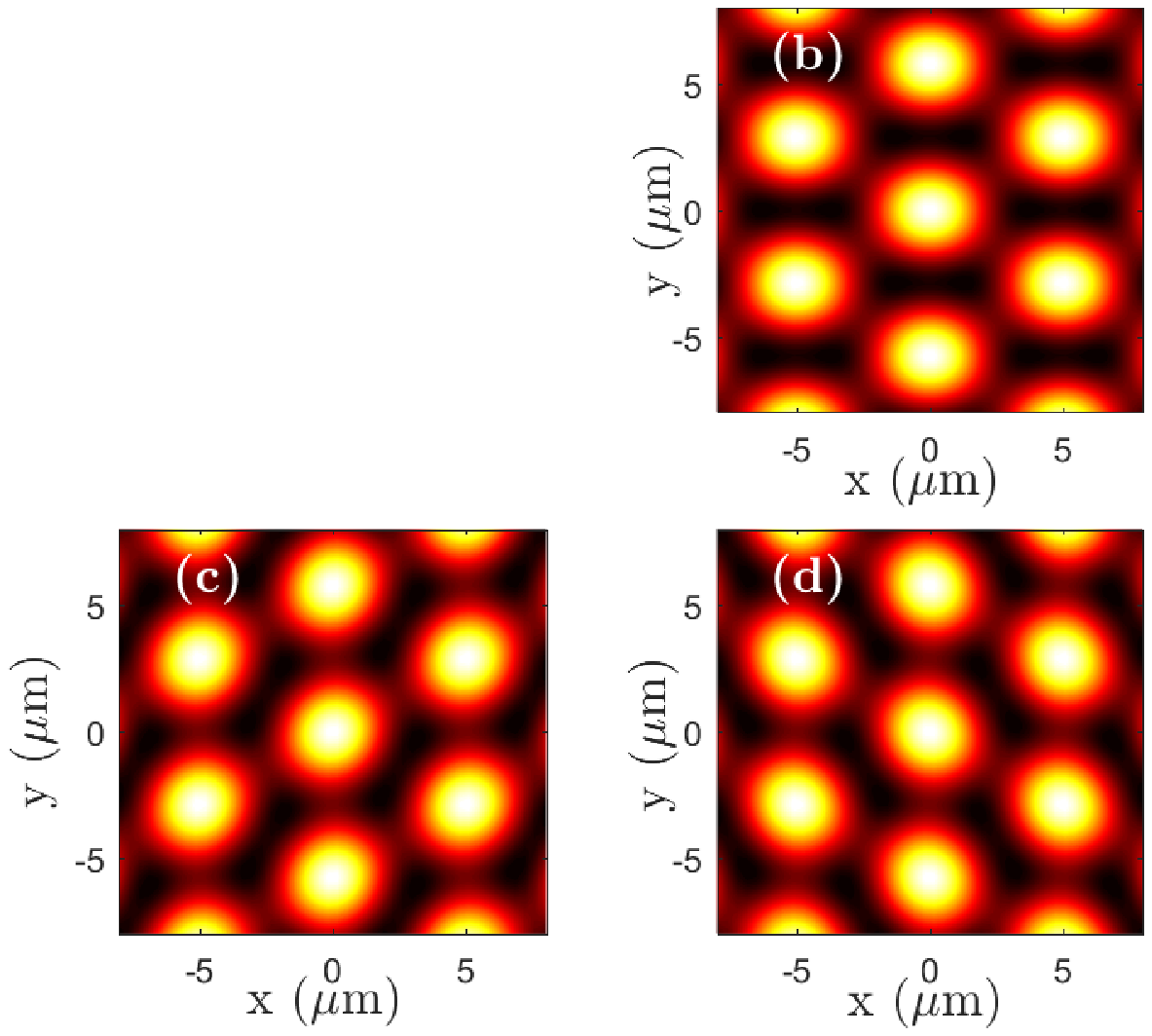}}
\includegraphics[trim=1.8cm 0.1cm 1.1cm .1cm, clip=true,height=8cm]{Fig1b.eps}\llap{\makebox[\wd1][l]{\raisebox{4.5cm}{
\begin{overpic}[height=3.2cm]{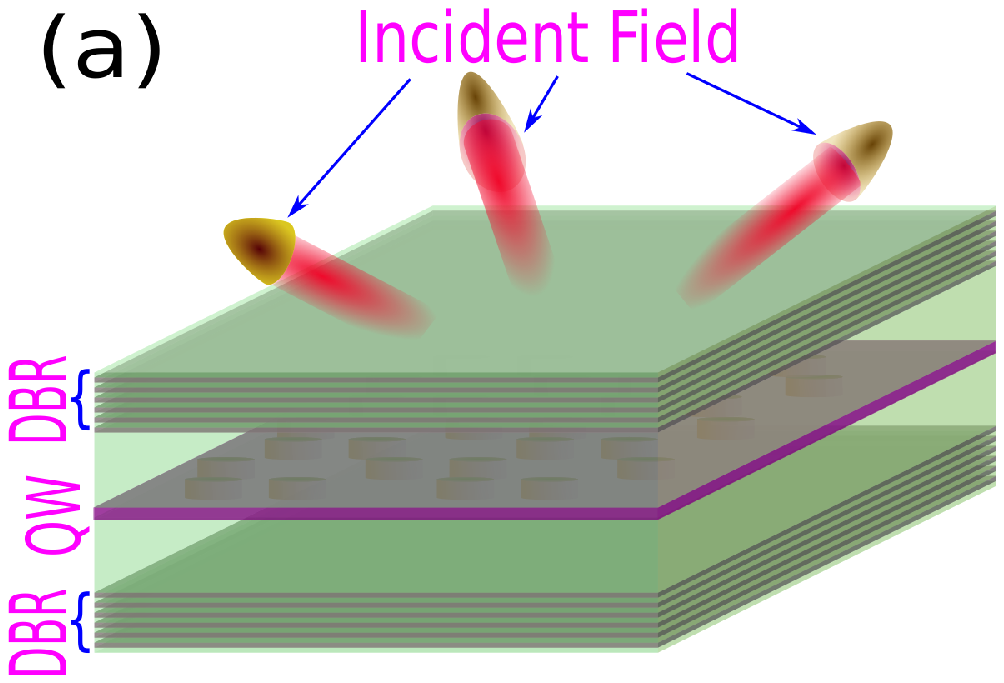}
\begin{footnotesize}
\put(2,-9){{\bf\color{mmagenta}DBR}: distributed bragg reflectors}
\put(25,-1){{\bf\color{mmagenta}QW}: quantum well}
\end{footnotesize}
\end{overpic}
}}}
\caption{Scheme of polaritons in a semiconductor microcavity with optical pumping fields used to create a time-dependent potential landscape. (b-c) show the potential at $t=0, 2\pi/3\omega, 4\pi/3\omega$. This is obtained from interfering two fields of the form $\exp(-i(\omega_0\mp\omega/2)t)\sum_m\exp(i{\bf k}_m\cdot({\bf r}_{2D}-{\bf r}_m)\pm\phi_m)$, where the wavevectors are $\mathbf{k_1}=(0,2\pi/\sqrt3a)$, $\mathbf{k}_{2,3}=(\mp\pi/a, -\pi/\sqrt3a)$; the phases are $\phi_1=0$, $\phi_{2,3}=\mp2/3\pi$, and the field displacements are $\mathbf{r_1}=(a/2,0)$, $\mathbf{r}_{2,3}=(\mp a/4,\sqrt3a/4)$. We consider the case $a=5~\mu$m, corresponding to a typical scale of polariton potentials.}
\label{f1}
\end{figure}

We consider a semiconductor microcavity containing quantum wells as illustrated in Fig.~\ref{f1}(a), with an additional spatio-temporal polariton potential:
\begin{equation}
V({\bf r},t)=-V_s({\bf r})-V_d({\bf r},t).
\label{eq:gamma}
\end{equation}
Here $V_s({\bf r})=V_0\cos\frac{2\pi}{3}\sum_{j< l}\cos\big(({\bf k}_j-{\bf k}_l)\cdot{\bf r}\big)$ is a static part of a honeycomb shaped potential, with $V_0$ its amplitude. This can be hard-engineered into the microcavity structure, for example by etching micropillar arrays~\cite{polaritonlattice1} or placing a metal surface pattern on top of the structure~\cite{polaritonlattice2,polaritonlattice3}. Alternatively, it has been shown that the polariton potential can be engineered by optical excitation~\cite{opticalpotential1}, where coherent excitation below the polariton resonance energy allows ultrafast Stark control of the polariton potential~\cite{opticalpotential2}. Since our considered potential can be decomposed into well-defined wavevector components, we assume it can be introduced through the interference of counter-propagating fields or the use of an appropriate (spatial light) modulator. The Stark induced switching of 
polaritons~\cite{opticalpotential2} was proposed
with a radiative broadening of 0.1meV. Recently, state-of-art experimental work has also revealed polariton radiative broadening on the order of 
a couple of micro electronvolts~\cite{lifetimeep}. As this is two to three orders of 
magnitude smaller than the other relevant energy scales in our problem, 
we consider here the physics of the Hermitian system.

Furthermore, since lasers interfering with slightly different frequencies give rise to a time-periodic interference pattern, we assume it is also possible to introduce a time-periodic component to the honeycomb potential, $V_d({\bf r},t)=\gamma V_0\sum_{l< j}^3\cos(\omega t-\phi_k)\cos\big({\bf k}_j\cdot({\bf r}-{\bf r}_j)-{\bf k}_l\cdot({\bf r}-{\bf r}_l)\big)$ as is shown in Fig.~\ref{f1}(a) for two lasers with detuning $\omega$, where $\gamma$ is a parameter defining the relative strength compared to the static potential. In the following analysis we fix $\gamma=1/10$ for simplicity.

This form of time-dependent potential is further illustrated in Figs.~\ref{f1}(c-d). Dynamic potential lattices can also be generated by interfering surface acoustic waves in a similar way~\cite{surfwave1,surfwave2}. Given that it breaks time reversal symmetry, which is required together with the parity symmetry to protect the Dirac cone for the 2D structure, we can expect a gap to be opened at the Dirac point. For a two-band model, this would drive the system into a topological insulator phase with $\mathcal{C} = \pm1$ in the large frequency limit. However, realistically, the frequency should be carefully chosen to make sure the system would not get heated too fast~\cite{HFE1}. In the following we will work in the opposite regime of the large frequency limit to (1) keep the system cooling down in the whole process and to (2) explore the more exotic physics in the regime where there is significant coupling between Floquet bands.

\subsection{Continuous model}

To make a full and reliable analysis we start with a continuous model of the potential, where the system is described by the Schr\"{o}dinger equation:
\begin{align}
i\hbar\frac{\partial\Psi(x,y,t)}{\partial t} = \bigg(-\frac{\hbar^2\nabla^2}{2m_{\rm eff}} + V(x,y,t)\bigg)\Psi(x,y,t).\label{eq:Shrodinger}
\end{align}
For the potential with period $T$, the long time behavior is given by the Floquet operator~\cite{HFE1},
\begin{align}
U(T) = \mathcal{T}\exp\bigg(-\frac{i}{\hbar}\int_0^T H(t)dt\bigg).
\end{align}
For our analysis we choose the frequency $\omega$ smaller than the bandgap between the second and third bands of the continuous Hamiltonian $H(t)$. 
The quasienergy spectrum is $\varepsilon_{\rm quasi}({\bf k}_B) = -{\rm imag}\{\hbar\log(\mathcal{E}[U_{{\bf k}_B}(T)])\}$, where ${\rm imag}$ represents the imaginary part and $\mathcal{E}[U_{\bf k_B}(T)]$ represents the eigenvalues of $U_{{\bf k}_B}(T)$. For the numerical calculation we use the plane wave expansion method: starting from $t=0$, we obtain the spectra of the first two bands of $H(t=0)$, $\{\varepsilon_{1/2}({\bf k}_B),\,|\psi_{1/2}({\bf k}_B)\rangle\}$, and the projector operator $\mathcal{P}_{{\bf k}_B} = |\psi_1({\bf k}_B)\rangle\langle\psi_1({\bf k}_B)|+|\psi_2({\bf k}_B)\rangle\langle\psi_2({\bf k}_B)|$. The full evolution operator $U_{{\bf k}_B}$ is obtained as $\langle {\bf k}_i |U_{{\bf k}_B}|{\bf k}_j\rangle = \langle {\bf k}_i|\mathcal{T}\exp(-\frac{i}{\hbar}\int_0^TH(t)dt)|{\bf k}_j\rangle\sum_{n_1,n_2}\delta_{{\bf k}_B,n_1{\bf G}_1+n_2{\bf G}_2+{\bf k}_j}$ where $n_i$ are integers, $i=1,2$, and ${\bf G}_i$,  are the reciprocal lattice parameters. Then, the full operator is projected into the Hilbert space spanned by the first two bands, $U^P_{{\bf k}_B} = \mathcal{P}_{{\bf k}_B}U_{{\bf k}_B}\mathcal{P}_{{\bf k}_B}$. If the projected space is decoupled from the rest of the Hilbert space, then $U_{{\bf k}_B}^P$ will be a real unitary operator, with eigenvalues distributed on the unit circle of the 2D complex plane. Otherwise, the modulus of the eigenvalues will deviate from the unit circle, for initial population in the spanned space, which will induce the decay of the population.

\begin{figure}
\centering\includegraphics[trim=1.8cm 0.1cm 1.1cm .1cm, clip=true,width=.99\columnwidth]{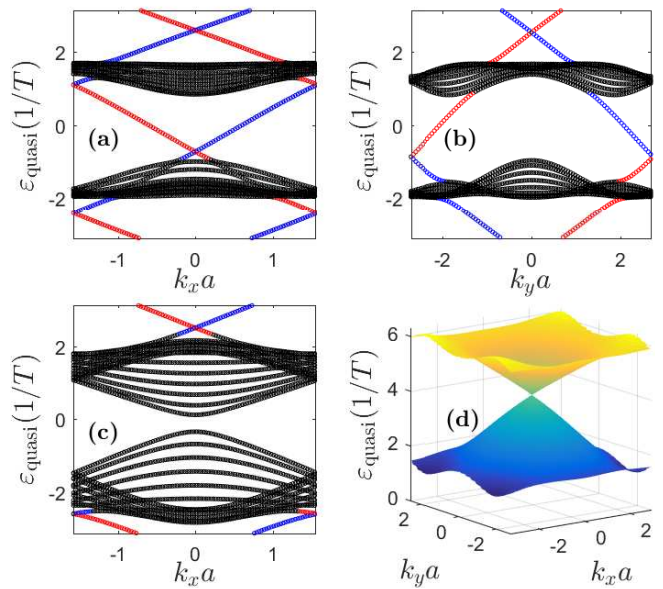}
\caption{Quasienergy dispersion of Floquet topological insulator with armchair (a) and zigzag (b) boundaries($V_0 = 1.50~$meV, $\omega = 0.20~$meV); (c) Chern insulator with armchair boundary ($V_0 = 1.80~$meV, $\omega=0.32~$meV). (d) Single Dirac cone at the critical point (from topological insulator to Floquet topological insulator) when $\omega$ decreases from infinity ($V_0 = 1.80~$meV, $\omega\approx30.84\times10^{-2}~$meV). Red colors mark edge states on the left/bottom boundaries for the armchair and zigzag structures, respectively; blue colors mark edge states for right/top boundaries.}
\label{f2}
\end{figure}

The quasi-bandstructure is shown in Figs.~\ref{f2}(a-b) for armchair and zigzag boundaries, respectively. Unlike the ordinary Chern insulator, there are two sets of edge states on each of the edges with one coming out of the top of the higher band and ending at the bottom of lower band, while the other set bridges directly the higher and lower bands in the Floquet zone. The Chern numbers of both bands calculated with the 2D structure are $\mathcal{C} = 0$, which indicates that the state observed here is the Floquet topological state; and both of the anomalous edges are topologically protected.

\subsection{Tight binding model}

It is instructive to consider a two-band tight-binding description. As is shown in Figs.~\ref{f1}(b-d) the whole period of evolution could be separated into three stages ($n=1,2,3$), each modelled with a different effective tight-binding Hamiltonian:
\begin{equation}
H_n({\bf k_B}) = \left(\begin{array}{cc}0&J_n({\bf k_B})\\J_n({\bf k_B})^*&0\end{array}\right).
\end{equation}
In the first stage, the coupling is stronger in the horizontal direction ${\bf d}_1 = (2a/3,0)$, rather than the directions of the other lattice sites ${\bf d}_{2/3} = (\pm a/3, a/\sqrt{3})$. Consequently, we write the coupling between bands as:
\begin{equation}
J_1({\bf k_B})=J\left(\alpha e^{-i{\bf k}_B\cdot{\bf d}_1}+\beta e^{-i{\bf k}_B\cdot{\bf d}_2}+\beta e^{-i{\bf k}_B\cdot{\bf d}_3}\right)
\end{equation}
where $J$ describes the overall coupling strength; $\alpha>1$ and $\beta<1$ are parameters determining the relative contributions of coupling in the different directions. Similarly:
\begin{align}
J_2({\bf k_B})&=J\left(\beta e^{-i{\bf k}_B\cdot{\bf d}_1}+\alpha e^{-i{\bf k}_B\cdot{\bf d}_2}+\beta e^{-i{\bf k}_B\cdot{\bf d}_3}\right)\\
J_3({\bf k_B})&=J\left(\beta e^{-i{\bf k}_B\cdot{\bf d}_1}+\beta e^{-i{\bf k}_B\cdot{\bf d}_2}+\alpha e^{-i{\bf k}_B\cdot{\bf d}_3}\right)
\end{align}
and the Floquet operator is given by:
\begin{align}
U_{tb}({\bf k}_B) = e^{-iH_3({\bf k_B})T/3}e^{-iH_2({\bf k_B})T/3}e^{-iH_1({\bf k_B})T/3}.
\end{align}
The parameter $J$ can be obtained from fitting the full bandwidth of the first two bands of the continuous model with $\gamma=0$ and $\alpha=\beta=1$. Then, $\alpha$ and $\beta$ are fixed by fitting the two band gaps of the quasi-energy bands at ${\bf k}_B=0$ for the continuous model with armchair boundary. Such a fit is shown in the Appendix and gives
the same physics of anomalous edges states (with zero Chern number) as expected.

\section{result and discussion}
\label{S:result}
\subsection{Chern insulator and Floquet topological insulator}

By increasing the depth of the potential and/or increasing the frequency of the time dependent potential, we could obtain a Chern insulator (with Chern number $\mathcal{C}=\pm1$), as shown in Fig.~\ref{f2}(c) and the same appears in the tight-binding model with different tight-binding parameters. The topological phase transition from Chern insulator to Floquet topological insulator is observed when the frequency of the potential is decreased from infinity to the order of the energy scale of the Hamiltonian system. As is shown in Fig.~\ref{f2}(d), at this transition point there is a single Dirac point in the Brillouin zone, which is strange according to symmetry analysis but comes from the fact that the different Floquet branches touch each other. In time reversal symmetry breaking systems, an unpaired fine-tuned Dirac point at the critical point could survive the ``fermion doubling" theorem, and this has been observed in both static~\cite{topo2,pan3} and Floquet~\cite{ftopola,photfto2} systems. This is the case when there is no degeneracy at both the top and bottom of the relevant bands for direct band material, where they would touch each other at the critical point when the time dependent period is approaching the full relevant bandwidth of the system beginning from infinity; in principle, for indirect band gap material different effects could occur.

\begin{figure}
\centering\includegraphics[trim=0.8cm 0.1cm 1.1cm 3.1cm, clip=true,width=.99\columnwidth]{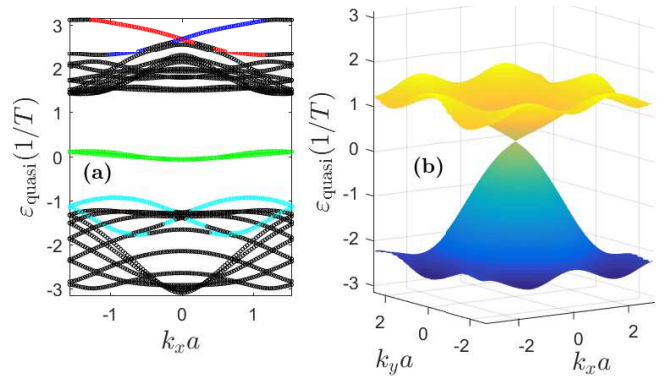}
\caption{Quasienergy dispersion for armchair structure. (a) Chern insulator, $\mathcal{C}=\pm1$, with $\omega=0.12~$meV; red/blue represent topologically protected edge states at left/right boundaries respectively; green indicates the flat band formed by edge states; cyan indicates the third piece of edge states. (b) the second phase transition point around $\omega=15.98\times 10^{-2}~$meV. Other parameters: $V_0 = 1.50~$meV. }
\label{f3}
\end{figure}

As the frequency $\omega$ decreases even further with respect to the full relevant bandwidth of the system, another topological phase transition appears, and the system goes back into the Chern insulator phase from the Floquet topological insulator phase as shown in Fig.~\ref{f3}(a). Bulk calculation shows that the Chern number is $\mathcal{C}=\pm1$, but as is shown in Fig.~\ref{f3}(a) in green, an exotic flat band is observed. In contrast to earlier demonstrations of exciton-polariton flat-bands~\cite{flatband}, this flat band develops from a pair of edge states. Since this flat band is located away from the bulk modes, it could in principle be exploited as a slow mode for light to enhance nonlinear effects. In addition, there is a third pair of edge states which is pushed into the bulk bands as shown in cyan. The two-band tight-binding model was also able to reproduce this physics. It also deserves to be pointed out that it may be possible to obtain high Chern number insulators by exploiting further topological phase transitions; the tight-binding calculation shows that a Chern insulator with $C=\pm5$ could be obtained.

Figure~\ref{f3}(b) shows the band structure at the second phase transition point from the Floquet topological insulator to the Chern insulator as $\omega$ decreases further. Again, an unpaired Dirac cone is observed at this time.

\subsection{Spin degree of freedom}

Accounting for the spin-degree of freedom of polaritons, one should write two copies of the Schr\"{o}dinger equation (Eq.~\ref{eq:Shrodinger}), for the two spin components of the polariton wavefunction $\Psi_\pm$. The two spins
are typically coupled by the effects of spin-orbit coupling (TE-TM splitting), described by the Hamiltonian~\cite{TETMsplitting}
\begin{equation}
\mathcal{H}_\mathrm{TE-TM}=\Delta_{so}\left(\begin{array}{cc}0&\left(i\frac{\partial}{\partial x}+\frac{\partial}{\partial y}\right)^2\\\left(i\frac{\partial}{\partial x}-\frac{\partial}{\partial y}\right)^2&0\end{array}\right)\label{eq:SC}
\end{equation}
where $\Delta_{so}$ determines the strength of spin-orbit coupling. \\

\begin{figure}
\centering\includegraphics[ clip=true,width=.8\columnwidth]{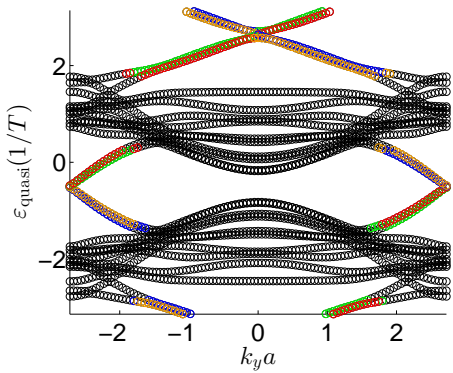}
\caption{Quasienergy dispersion, $\varepsilon_{\rm quasi}$ including spin dependence for the zigzag structure. $V_0 = 1.5~$meV, $\omega=0.30~$meV and $\Delta_{so} = 0.05~$meV$\mu$m$^2$. Red/Blue indicate the modes on the left/right edges, respectively, for the $+$state. Orange/green indicate the modes on the left/right edges for the $-$state. }
\label{f4}
\end{figure}

Fig.~\ref{f4} shows the result of a tight binding calculation of the band structure in the zigzag edge configuration. The propagation direction of the topologically protected edge states depends on the polariton spin state: for the $+$state the mode on the left edge (in red) moves in the $+y$ direction, on the right edge (in blue) it moves in the $-y$ direction. For the $-$state, the mode on the left edge (in orange) moves in the $-y$ direction, and the one on the right edge (in green) moves in the $+y$ direction. This behavior is due to a choice of polarization dependent potential, $V_d$, which may have opposite signs for different spin polarizations if an appropriate polarized laser field is used to excite the system. The phenomena where the two polarizations behave similarly is also possible if $V_d$ takes the same sign for different polarizations.

\section{summary}
\label{S:3}

We have proposed to achieve Floquet topological polaritons, based on the time-dependent potential modulation of a semiconductor microcavity, which could be achieved optically. By working in the regime where the Floquet frequency is comparable with the relevant bandwidth of exciton-polaritons, we showed that a Chern insulator can be obtained. By decreasing the Floquet frequency we came across the first critical point where an unpaired Dirac cone was observed, and the Floquet topological insulator was obtained with $\mathcal{C}=0$. By further decreasing the Floquet frequency, the second phase transition was observed with unpaired Dirac cone, and an exotic Chern insulator with flat band was obtained. Finally, considering the spin-degree of freedom of polaritons allows for polaritons with different spin polarizations to either co-propagate or counter-propagate in topological edge modes, depending on the choice of polarized optical potential. The latter case would correspond to a spin-momentum locking of polaritons, suitable for the generation and filtering of spin currents. In future work it would be interesting to study the interplay between the obtained topological states and nonlinear interactions, which could allow further control of topological states~\cite{topononlinear1,topononlinear2} and the formation of solitons useful for information processing~\cite{topopolar5,soliton1,soliton2,soliton3}.

This research was supported by the Ministry of Education (Singapore) grant 2015-T2-1-055.\\

\appendix*
\section{Tight-binding calculation}
For convenience, we will briefly repeat the physics of the tight-binding model. As shown in Figs.1(b-d) in the main text, the whole process could be separated into three stages.

Stage one:
\begin{equation} 
H_{{\bf k}_B}^1 =  \sum_jH_{{\bf k}_B}^{1}({\bf d}_{j}),
\end{equation} 
with
\begin{align}
H_{{\bf k}_B}^1({\bf d}_1) = \alpha J\begin{pmatrix}
0 & e^{-i{\bf k}_B\cdot{\bf d}_1} \\
e^{i{\bf k}_B\cdot{\bf d}_1} & 0
\end{pmatrix},
\end{align}
and
\begin{align}
H_{{\bf k}_B}^{1}({\bf d}_{2/3}) = \beta J\begin{pmatrix}
0 & e^{-i{\bf k}_B\cdot{\bf d}_{2/3}} \\
e^{i{\bf k}_B\cdot{\bf d}_{2/3}} & 0
\end{pmatrix}.
\end{align}

Stage two: 
\begin{equation}
H_{{\bf k}_B}^2 =  \sum_jH_{{\bf k}_B}^{2}({\bf d}_{j}),
\end{equation} 
with 
\begin{align}
H_{{\bf k}_B}^2({\bf d}_2) = \alpha J\begin{pmatrix}
0 & e^{-i{\bf k}_B\cdot{\bf d}_2} \\
e^{i{\bf k}_B\cdot{\bf d}_2} & 0
\end{pmatrix},
\end{align}
and
\begin{align}
H_{{\bf k}_B}^{2}({\bf d}_{1/3}) = \beta J\begin{pmatrix}
0 & e^{-i{\bf k}_B\cdot{\bf d}_{1/3}} \\
e^{i{\bf k}_B\cdot{\bf d}_{1/3}} & 0
\end{pmatrix}.
\end{align}

Stage three: 
\begin{equation}
H_{{\bf k}_B}^3 =  \sum_jH_{{\bf k}_B}^{3}({\bf d}_{j}),
\end{equation} 
with 
\begin{align}
H_{{\bf k}_B}^3({\bf d}_3) = \alpha J\begin{pmatrix}
0 & e^{-i{\bf k}_B\cdot{\bf d}_3} \\
e^{i{\bf k}_B\cdot{\bf d}_3} & 0
\end{pmatrix},
\end{align}
and
\begin{align}
H_{{\bf k}_B}^{3}({\bf d}_{1/2}) = \beta J\begin{pmatrix}
0 & e^{-i{\bf k}_B\cdot{\bf d}_{1/2}} \\
e^{i{\bf k}_B\cdot{\bf d}_{1/2}} & 0
\end{pmatrix}.
\end{align}

\begin{figure}
\centering\includegraphics[trim=1.8cm 0.1cm 1.1cm .1cm, clip=true,width=.99\columnwidth]{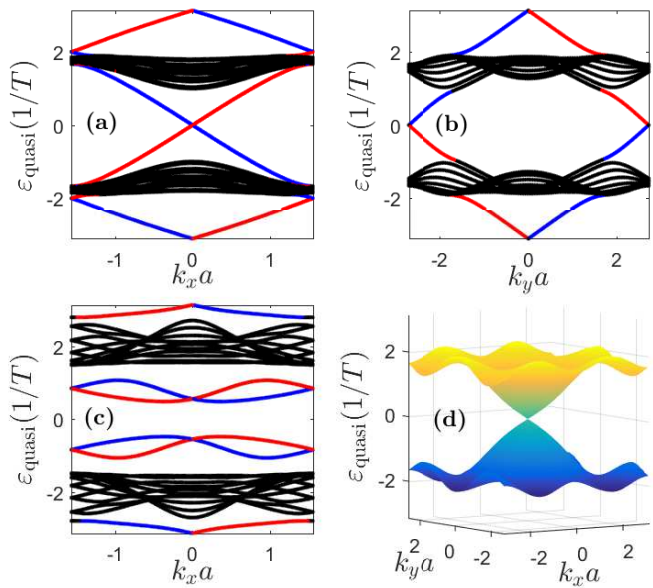}
\caption{Tight-binding calculation of quasi-spectra, $\varepsilon_{\rm quasi}$ for potential strength $V_0 = 1.50~$meV with $\gamma = 0.10$. (a)/(b) are Floquet topological insulators with armchair and zigzag boundary conditions respectively ($\omega = 0.20~$meV); (c) Chern insulator with armchair boundary condition ($\omega=0.12~$meV). (d) Single Dirac cone at the phase transition point around $\omega_{c_2} \approx 16.04\times 10^{-2}~$meV (from Floquet topological insulator to Chern insulator) when $\omega$ decreases. Red colors are for edge states on the left/bottom boundaries for the armchair and zigzag structure respectively, and blue for right/top boundaries.}
\label{fs1}
\end{figure}

For potential strength $V_0 = 1.50~$meV and $\gamma=0.10$, we have the relevant parameters, $J = 3.80\times 10^{-2}~$meV and $\alpha = 3.30$, $\beta = 0.46$ and the quasi-energy structure calculated from the Floquet operator $U_{tb}({\bf k}_B) = e^{-iH_{{\bf k}_B}^3T/3}e^{-iH_{{\bf k}_B}^1T/3}e^{-iH_{{\bf k}_B}^1T/3}$, is shown in Fig.~\ref{fs1}. For $\omega = 0.20~$meV, we get the Floquet topological insulator~\cite{ftopochar1,ftopochar2}, with two sets of the edge states on each of the edges, which is consistent with the continuous calculation (and the Chern number~\cite{Cherncalculation} is zero). Around $\omega \approx 16.04\times10^{-2}~$meV, the topological phase transition with unpaired Dirac cone is observed as is shown in Fig.~\ref{fs1}(d); all these calculations are consistent with the continuous picture. At $\omega=12.00\times10^{-2}~$meV, the system is an exotic Chern insulator, $\mathcal{C}=\pm1$ for which the tight-binding calculation could only qualitatively get the relevant physics. As is shown in Fig.~\ref{fs1}(c), there are two strange edge states in the bulk gap; from the full continuous calculation, we know that one set of the states is squeezed into a flat band while the other set is pushed into the bulk spectra.

\begin{figure}
\centering\includegraphics[trim=1.8cm 0.1cm 1.1cm .1cm, clip=true,width=.99\columnwidth]{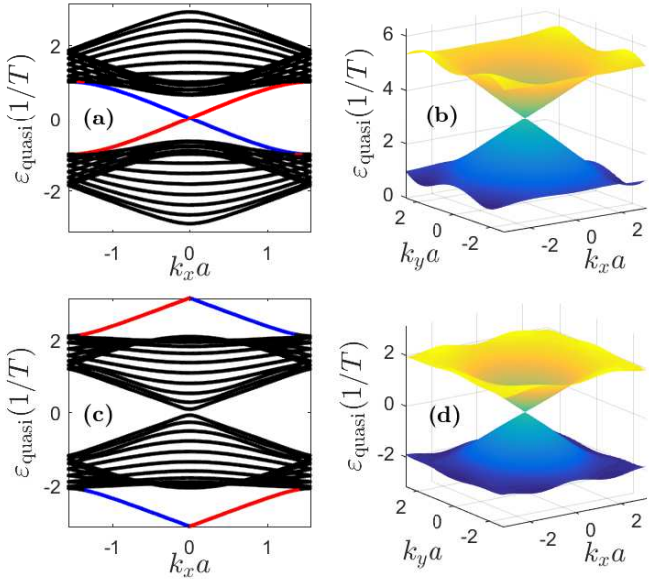}
\caption{Tight-binding calculation of quasi-spectra, $\varepsilon_{\rm quasi}$ for armchair structure with potential strength $V_0 = 1.80~$meV, $\gamma = 0.10$. (a) Chern insulator ($\omega = 0.32~$meV); (b) Single Dirac cone at the first phase transition point $\omega_{c_1} \approx 30.74\times 10^{-2}~$meV (from Floquet topological insulator to Chern insulator) when $\omega$ decreases. (c)/(d) Topologically protected edge state/unpaired Dirac cone at the second phase transition point $\omega_{c_2} \approx 15.38\times 10^{-2}~$meV as $\omega$ decreases further. The same color scheme is used as in Fig.~\ref{fs1}.}
\label{fs2}
\end{figure}

Figure~\ref{fs2} shows the calculation for $V_0=1.80~$meV, $\gamma=0.10$. The relevant parameters are $J = 3.15~\times10^{-2}~$meV, $\alpha=4.48$ and $\beta = 0.20$. For $\omega=0.32~$meV, a Chern insulator, $\mathcal{C}=\pm1$, is obtained as in Fig.~\ref{fs2}(a); and the first phase transition happens around $\omega_{c_1}\approx 30.74\times10^{-2}~$meV (Fig.~\ref{fs2}(b)). Around $\omega_{c_2}\approx 15.38\times10^{-2}~$meV the second phase transition is observed with unpaired Dirac cone and one set of topologically protected edge state as shown in Fig.~\ref{fs2}(c-d).

\end{document}